\documentstyle[aps]{revtex}
\def \be   {\begin{equation}}
\def \ee   {\end{equation}}
\def \l {\label}
\begin{document}
\input epsf
\baselineskip=25pt
\title{Conformally symmetric massive discrete fields}
\author{Manoelito M de Souza\footnote{Permanent address:Departamento de
F\'{\i}sica - Universidade Federal do Esp\'{\i}rito Santo\\29065.900 -Vit\'oria-ES-Brazil- E-mail: manoelit@cce.ufes.br}}
\address{Centro Brasileiro de Pesquisas F\'{\i}sicas- CBPF\\R. Dr. Xavier Sigaud 150,\\
22290-180 Rio de Janeiro -RJ - Brazil}
\date{\today}
\maketitle
\begin{abstract}
\noindent Conformal symmetry is taken as an attribute of theories of massless fields in manifolds with specific dimensions. This paper shows that this is 
not an absolute truth; it is a consequence of the mathematical representation used for the physical interactions. It introduces a new kind of representation 
where the propagation of massive (invariant mass) and massless interactions are unifiedly described by a single conformally symmetric Green's function. 
Sources and fields are treated at a same footing, symmetrically, as discrete fields - the fields in this new representation - fields defined with support on 
straight lines embedded in a (3+1)-Minkowski manifold. 
The discrete field turns out to be a  point in phase space. It is finite everywhere. With a finite number of degrees of freedom it does not share the well 
known problems faced by the standard continuous formalism which can be retrieved from the discrete one by an integration over a hypersurface. The passage 
from discrete to continuous fields illuminates the physical meaning and origins of their properties and problems. The price for having massive discrete 
field with conformal symmetry is of hiding its mass and timelike velocity behind its non-constant proper-time. 
\end{abstract}
\begin{center}
\end{center}
\section{Introduction}

A cone can be seen as a continuous
set of straight lines (the cone generators) intersecting on a single point
(the cone vertex). This simple heuristic picture is the basis for introducing a new
tool in field theory, the concept of a discrete field. Technically, it is a field
defined with support  on a straight line embedded in a $(3+1)$ Minkowski
spacetime instead of the usual lightcone support of the radiation fields. More pictorially this is a discretization of the standard continuous field when 
one sees it as a set of points and deals with each one instead of with the entire set as a whole. It carries a continuous label indicative of its 
lightcone-generator support so that an integration over this label reproduces the usual (massive and massless) continuous fields and their complete 
formalisms. A single Green's function, symmetric under conformal transformations, works as the propagator of both massive and massless discrete fields. This 
function, as well as the wave equation and the Lagrangian, cannot be explicit functions of the mass parameter and  this would break their conformal symmetry. The masses reappear in the formalism (in the Lagrangian, in the wave equation and in the Green's function) when the continuous formalism is retrieved with an integration over the discrete fields.  

The point on introducing the discrete field is that it does not have the problems of the continuous ones. In particular it is a finite field, free of 
problems with infinities, causality violation and spurious degrees of freedom. It is just a point in phase space, symmetric under conformal transformations,
univocally determined by its source and propagating with a well defined and everywhere conserved energy-momentum content. So it can be treated as a 
legitimate point-like physical object with a finite number of degrees of freedom, and this explains why it does not share the same problems of the 
continuous field with its infinite number of degrees of freedom. It is a very peculiar object with simultaneous characteristics of both a field and a 
particle; it may be a key step for a better comprehension of quantum theory.  These nice properties disappear after the integration that turns it into a 
continuous field.

Studying the transformation from the discrete to the continuous
field one can have a deeper understanding of the physical meaning and origin
of many of their properties and problems. Most of them
comes from their hypercone support; their singularity, for example, as will be shown,
is just a reflex of the hypercone vertex (Section V). This approach came from a study of classical electromagnetic radiation on
its limiting zero distance from its sources \cite{hep-th/9610028} and its problems of
infinite energy and causality violation \cite{Rorhlich,Jackson,Ternov,Teitelboim,Rowe,Lozada}. Its simplification power is drastically exhibited in
an application to the general theory of relativity \cite{gr-qc/9801040}: the
highly non-linear field equations are reduced, without any approximation, to
the wave equation in a (3+1) Minkowski spacetime, and yet from their discrete
solutions one can, in principle, retrieve any continuous solution from the full equation \cite{paperII,gr-qc/9903071}.
The simple change of support in the field definition has an immediate dynamical 
consequence valid for all fundamental interactions\cite{hep-th/9708096}: in the source instantaneous rest frame
 the field is always emitted along a direction
orthogonal to its source acceleration (Section IV). For the electromagnetic interaction, in particular, this constraint has a solid experimental confirmation \cite{Jackson,Ternov} that
validates its implementation as a basic physical input. This, however, will be discussed only in the companion paper III \cite{paperIII} as an application to the electromagnetic field, but part of it has been anticipated in the Section VIII of \cite{hep-th/9911233}.

 As it describes a (1+1) dynamics endowed with a conformal symmetry and  covariantly embedded in a (3+1)-spacetime, this approach has a relevance to all bi-dimensional field theory formalism \cite{Berkov}. A discrete field can be thought, in a string theory context \cite{string}, as a string in its zero-length limit. The good results in (1+1) statistical mechanics and field theory are extendable to (3+1) physics of discrete, massive and massless fields. This conformal symmetry of massive fields, it is necessary to emphasize, is not in the sense of introducing mass transformations \cite{Schouten,Barut}. We are referring to constant, invariant masses. Moreover, this is not a Kaluza-Klein formalism \cite{KK} although use is made, just for convenience, of a fifth (but timelike) dimension and so they have many points of contact.  It is indeed closer, despite its distinct goals and results to the seminal work of Dirac \cite{Dirac2times} and a more recent work done on this line \cite{Bars}, as both work with a projective geometry introducing a manifold with a second time dimension. Distinct aspects here are that there is no extra space-dimension and that the second time is a Lorentz scalar with the fixed meaning of a proper time. The use of a scalar second time is very frequent in the literature (see \cite{tau} and the references therein), with various qualifications (like the invariant, the universal, the historic time, etc), interpretations and goals.  Here it is the length of an interval associated to the propagation of a physical point-like object. Although we are restricting the subject to a classical treatment the discrete field is  defined in such a way that it can be extended to a quantum context too. This allows it to be relevant to the problems of field quantization; to quantum gravity, particularly. Discrete gravity of general relativity is discussed in the companion paper II \cite{paperII} as the only possible interpretation of a discrete scalar field.

We will discuss here the properties of a discrete-field Green's function which do not depend on the field tensorial nature. The unfeasibility of a strictly point-like physical signal will make the transition to a quantum context (not discussed here) mandatory but perhaps easier and more natural. Classical physics is then an idealized limit of the quantum one where point signals can be produced and measured.  But a proper discussion on physical meanings will be done with respect to specific fields on the companion papers II (on a classical scalar field\cite{paperII}) and III (on the vector field, or more properly, the Maxwell field of classical electrodynamics\cite{paperIII}). This separation is convenient for a gradual construction of a discrete field formalism and also for making explicit which properties are and which are not consequences of the spinor or tensor character of the field. Our strategy is of exhausting first the simplest idealized classical limit before making the necessary transition to the quantum context. 

This paper is organized in the following way. In Section II a new
geometrical way of implementing relativistic
causality in field theory is introduced. It extends the concept of local causality to make
possible a consistent and manifestly conformal covariant definition of discrete fields.  Discrete fields are properly defined in
Section III. A more direct and mathematically oriented approach (skipping this causality motivation) is presented in Section II of \cite{paperII}. Here we rather emphasize a more physical approach so that an intuitive feeling can be developed. Its implications to the field
dynamics and to the consistency of classical electrodynamics are discussed in Section IV. In Section V we find and discuss the properties of discrete-field solutions to the wave equations; how the propagation of both massive and massless fields in (3+1)-dimensions is unifiedly described by a same Green's function, whose conformal symmetry is proved in Section VI. In Section VII we discuss the relationship between the discrete and the continuous
field formalisms, how the continuous field and its wave equation are both retrieved from their respective discrete ones. In Section VIII we clarify how a discrete field can be simultaneously massive and conformally (chirally too, if fermionic, although not discussed here) symmetric. The paper ends with some
final comments and the conclusions in Section IX.
\section{ Local and extended causality} 

\noindent We recur to causality as a physical motivation for defining the discrete field in a consistent and manifestly covariant way. Any given pair of events on Minkowski spacetime defines a four-vector $\Delta x.$ If a  $\Delta x$ is connected to the propagation of a free physical object (a signal, a particle, a field, etc) it is constrained to satisfy
\be
\label{1}
\Delta\tau^2=-\Delta x^{2}, 
\ee 
where $\tau$ is a real-valued parameter. We use a metric $\eta=diag(1,1,1,-1)$. $\Delta\tau$ is the invariant length or norm of $\Delta x$. So, the 
constraint (\ref{1}) just expresses that $\Delta x$ cannot be spacelike. A physical object does not propagate over a spacelike $\Delta x.$ This is 
{\it local causality}.  Geometrically it is the definition of a three-dimensional double hypercone; $\Delta x$ is the four-vector separation between 
a generic event $x^{\mu}\equiv({\vec x},t)$ and the hypercone vertex. See the Figure 1.

\vglue-2cm 
\begin{minipage}[]{5.0cm}\hglue-3.50cm
\parbox[]{5.0cm}{
\begin{figure}
\epsfxsize=400pt
\epsfbox{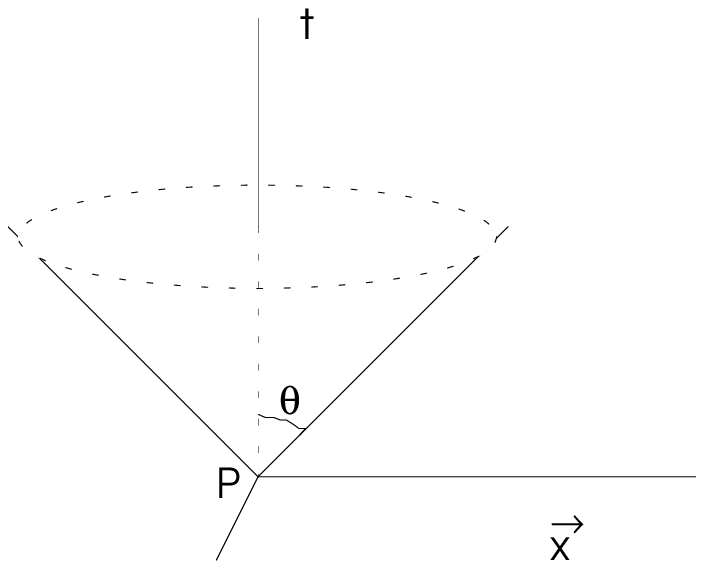}
\end{figure}}
\end{minipage}\hfill
\vglue-4cm
\vglue-12cm
\hglue7.0cm
\begin{minipage}[]{7.0cm}\hglue7.0cm
\begin{figure}
\l{f1}
\parbox[b]{7.0cm}
{\vglue-7cm \parbox[t]{7.0cm}{\vglue-1cm
\caption[Local causality]{The relation $\Delta\tau^2=-\Delta x^{2},$ a causality constraint, is seen as a restriction of access to regions of spacetime. It 
defines a three-dimension hypercone which is the spacetime available to a free physical object at the hypercone vertex. The object is constrained to be on 
the hypercone.}}}
\end{figure}
\end{minipage}\\ \mbox{} 

Changes of proper time are defined through intervals associated to the propagation of a free field in Minkowski spacetime and not through their association 
to trajectories as it is  usually done. This subtlety avoids a restriction to classical contexts and allows its application to quantum physics too.

The validity of Eq. (\ref{1}) is conditioned to its application to a free field but this is not such a great limitation as it seems to be at a first look if 
we assume that at a fundamental level all interactions are discrete. It describes the free evolution of an interacting field between any two consecutive 
interaction events. 

Local causality is usually implemented in special relativity through the use of lightcones by requiring that massive and massless objects remain, 
respectively inside and on a lightcone.  Our way of implementing the same relativistic causality is of using hypercones (not necessarily lightcones) even 
for massive physical objects (the expression physical object is used here for not distinguishing between particles and fields) as a constraint on their 
propagation. 
In spacetime a field is defined on hypersurfaces: hyperplanes for newtonian fields, for example, and hypercones for relativistic fields. Think of a wave 
front, for example, and think of it as a continuous set of moving points, then each point of it is on a world line tangent to a generator of its 
instantaneous hypercone. 

 This conic hypersurface, in field theory, is the support for the propagation of a free field: the field  cannot be inside nor outside but only on the 
 hypercone. The hypercone-aperture angle $\theta$ is given by $$\tan\theta=\frac{|\Delta {\vec x}|}{|\Delta t|},\; c=1,$$ or equivalently, by 
\be 
\l{extrm}
\Delta\tau^{2}=(\Delta t)^{2}(1-\tan^{2}\theta).
\ee
The speed of propagation determines the hypercone aperture (rapidity). A change of the supporting hypercone corresponds to a change of speed of propagation 
and is an indication of interaction.
Special relativity restricts $\theta$ to the range $0\le\theta\le\frac{\pi}{4},$ which  corresponds to a restriction on 
$\Delta\tau:$ $0\le|\Delta\tau|\le|\Delta t|.$ The lightcone ($\theta=\frac{\pi}{4},$ or $|\Delta\tau|=0$) and the t-axis in the object rest-frame 
($\theta=0,$ or $|\Delta\tau|=|\Delta t|$) are the extremal cases: the lightcone for objects with the speed of light, and a parallel line to the time-axis 
for each point of an static field or of a massive object on its rest frame.

For defining a discrete field we will need a more restrictive constraint:
\be
\l{f}
\Delta\tau+f.\Delta x=0,\;\;\;{\hbox{for}}\;\;\Delta x\neq0,
\ee
where $f$ is defined by 
\be
\l{ffrac}
f^{\mu}:=\frac{\Delta x^{\mu}}{\Delta\tau},
\ee
 a constant  four-vector tangent to the hypercone; it is  timelike $(f^{2}=-1$) if $\Delta\tau\neq0,$  or a limiting lightike four-vector $(f^{2}=0$) if $\Delta\tau=0$ and $\Delta x\neq0$. Observe that $f^{\mu}$ is well defined, as a tangent vector to the lightcone, for $\Delta\tau=0$ and $\Delta x\neq0$, but not for $\Delta\tau=\Delta x=0$, as a tangent vector is not well defined at the cone vertex. This is connected to the consistency of classical electrodynamics in the zero-distance limit discussed in \cite{hep-th/9708096}.

The Eq. (\ref{f})  defines a hyperplane tangent to the hypercone (\ref{1}). Together, Eqs. (\ref{1}) and (\ref{f}) define a hypercone generator $f$, tangent to $f^{\mu}$. A fixed four-vector $f^{\mu}$ at a point labels a fibre in the spacetime, a straight line tangent to $f^{\mu}$, the $f$-generator of the local hypercone (\ref{1}).\\
On the other hand the Eq. (\ref{f}) also implies that
\be
\l{fmu}
f_{\mu}=-\frac{\partial\tau}{\partial x^{\mu}},
\ee
with  $f_{\mu}=\eta_{\mu\nu}f^{\nu}$ and $\tau$ seen then as a known function of $x$, given by the constraint (\ref{1}). For $\Delta\tau=0$, that is, for a massless field, $f_{\mu}$ is the four-vector normal to the tangent hyperplane (\ref{f}).

Extended causality is the imposition of both Eqs. (\ref{1}) and (\ref{f}) to the propagation of a  physical object. Geometrically, it is a requirement that the point object remains on the hypercone generator $f$. It is, of course, much more constraining than local causality, the imposition  of just the Eq. (\ref{1}). This corresponds to a change in our perception of the spacetime causal structure; instead of seeing it as a local foliation of hypercones (\ref{1}) we see it as  congruences of lines (Eqs. (\ref{1}) and (\ref{f}) together); instead of dealing with continuous and extended objects, like a (standard) field for example, we treat them  as sets of points (discrete fields). 

\section{Discrete fields}

The imposition of extended causality corresponds then to a discretization of a physical system and should be distinguished from the quantization process, if for nothing else because it can be applied to both classical and quantum systems. On the other hand, there exists a relation of complementarity between local and extended causality, in the same sense of the one existing between geometric and wave optics as descriptions of light in terms of wave fronts and rays, respectively. Extended causality is a natural description for a classical particle\cite{hep-th/9708096} but its application to a classical field makes it, the discrete classical field, the closest thing to the classical counterpart of the quantum of its respective quantum field. In this paper however, we have no intention of pre-assigning any physical interpretation to a discrete field. Let us work it, for now, as just a convenient tool. 

\noindent Let us turn now to the question of how to define, in a consistent and manifestly covariant way, a field with support on a generic fibre $f$, a $(1+1)$-manifold embedded on a $(3+1)$-Minkowski spacetime. 
Every physical  field is tied through the proper time to its source, or better, to its creation event. There is nothing new or special on this: a massless field, for example, propagates on the lightcone and so its proper time does not change; its clock keeps marking the time of its creation, the instantaneous proper-time of its source at the event of its emission. This is one of the extreme situations depicted in the paragraph right after Eq. (\ref{extrm}); the other one is that of a static field for which  $\Delta\tau=\Delta t$ at each point. 
We use this as a way of implementing causality \cite{gr-qc/9801040,hep-th/9610145,BJP}. With $\tau$  being a known function of $x$, a solution of Eq. (\ref{1}),  
\be
\tau=\tau_{0}\pm\sqrt{-(\Delta x)^{2}},
\ee
to write $\Phi(x)$ for a field is the same (or almost) as writing $\Phi(x,\tau(x)).$ The subtlety of including $\tau(x)$ is of encoding in the very field the causal constraint (\ref{1}) on its propagation. 
Then to say, for example, that $\Delta\tau=0$ for a field, is just an implicit way of requiring that it propagates on a lightcone.  In order to implement local causality for a given field $\Phi(x)$, therefore, we just have to insert in it an explicit dependence on $\tau(x)$. For example, the replacement
\be
\l{6p}
\Phi(x)\Longrightarrow \Phi(x-z,\tau(x)-\tau(z)){\Big |}_{{\tau(x)=\tau(z)}}
\ee
describes a radiation field on the lightcone propagating from an event $z$ to an event $x$. We will, in general, write just  
\be
\l{6pp}
\Phi(x)\Longrightarrow \Phi(x,\tau){\Big |}_{\tau=0},
\ee
 for short. A static field, following this line of thought, is just a particular case where  $$\Phi(x)\Longrightarrow \Phi(x,\tau){\Big |}_{{\tau=t}}$$ which is, naturally, a frame dependent expression, in contradistinction to the previous one.  
 
The four-vector potential of radiation fields in classical electrodynamics and, in particular, of the Lienard-Wiechert solution \cite{Rorhlich,Jackson,Ternov,Teitelboim} have support on the lightcone  and they are well known examples of how a propagating field depends on the proper time of its source. This dependence, let us repeat, is just a form of causality implementation.
 
Whereas the implementation of local causality requires a field defined with support on hypercones,  for extended causality it is required a field with support on a line $f$:
\be
\l{fitotfif}
\Phi(x)\Longrightarrow \Phi(x-z,\tau(x)-\tau(z)){\Big |}_{{\Delta\tau+f.\Delta x=0}\atop{\Delta\tau^2+\Delta x^2=0}}
\ee
Let $\Phi_{f}$ represent such a field  
$$\Phi_{f}(x-z,\tau_{x}-\tau_{z})=\Phi(x-z,\tau_{x}-\tau_{z}){{\Big |}_{{\Delta\tau+f.\Delta x=0}\atop{\Delta\tau^2+\Delta x^2=0}}}:=\Phi(x,\tau){\Big |}_{f},$$
with ${\Big |}_{f}$ being a short notation for the double constraint ${{\Big |}_{{\tau+f.x=0}\atop{\tau^2+x^2=0}}}.$  Again, for a lighter notation, sometimes we just omit $z$ and $\tau_{z}$. It is called a discrete field, for reasons to become clear later.

It would not make any sense defining such a field if this restriction (to a line) on its support could not be sustained during its time evolution governed by the standard wave equation in $(3+1)$-dimensions. It is remarkable, as we will see in Section V, that this makes a consistent field definition.

 \noindent The derivatives of $\Phi_{f}(x,\tau),$ allowed or induced by the constraint (\ref{f}), are the directional derivatives along $f,$ which with the use of Eq. (\ref{fmu}) we write as
\be
\label{fd}
(\frac{\partial }{\partial x^{\mu}}+\frac{\partial \tau}{\partial x^{\mu}}\frac{\partial}{\partial \tau})\Phi_{f}={\Big(}\frac{\partial }{\partial x^{\mu}}-f_{\mu}\frac{\partial}{\partial \tau}{\Big)}\Phi_{f}:=\nabla_{\mu} \Phi_{f},
\ee

With $\nabla$ replacing $\partial$ for taking care of the constraint (\ref{f}), $\tau$ can be treated as a fifth independent  coordinate, timelike and Lorentz invariant:
\begin{eqnarray}
\l{eq1}
\tau(x)\Rightarrow & x^{5}=\tau \\
\l{eq2}
\partial_{\mu}\Rightarrow &\nabla_{\mu}
\end{eqnarray}
 The constraint (\ref{1}) is used only afterwards then. We adopt this geometrical approach which corresponds to embedding the physical spacetime in a $(3+2)$-manifold, as discussed in \cite{gr-qc/9801040,BJP} and by replacing the Minkowski geometry by a projective one. The $f$ in the definition of $\nabla$ is specified by the constraints on the field. Let us illustrate:
\be
\nabla_{\mu}{\bigg(}A_{f}(x,\tau)+B_{f'}(x,\tau){\bigg)}=(\partial_{\mu}-f_{\mu}\partial_{\tau})A_{f}(x,\tau)+(\partial_{\mu}-{f'}_{\mu}\partial_{\tau})B_{f'}(x,\tau).
\ee

Although not widely recognized, the radiation electromagnetic field of classical electrodynamics, more specifically the Maxwell stress tensor, is a field explicitly defined \cite{Rorhlich,Jackson,Ternov,Teitelboim,Rowe} with the two constraints (\ref{1}) and (\ref{f}). So, geometrically it should be regarded as a field defined with support on a line $f$, but classical electrodynamics, in this respect, is not consistent because the Maxwell tensor is the curl of a vector field (the vector potential) defined with support on the lightcone. This is so because the significance of the constraint (\ref{f}) has not being fully recognized yet\cite{hep-th/9708096}. This becomes clear in the wave equation for the Maxwell tensor as it does not encompass the constraint (\ref{f}) as it should. This paper tries to develop a consistent (in this approach) geometrical treatment for a generic field of unspecified tensoriality/spinoriality. 

Let us consider a generic field equation represented by
\be
\l{gfeq}
{\cal O}(x,\eta,\partial,s)\Phi(x)=J(x),
\ee
where ${\cal O}(x,\eta,\partial,s)$ is a linear differential operator, $\eta$ the Minkowski metric tensor, and $s$ may represent other parameters or degrees of freedom like mass, spin, etc. The tensorial character of the field $\Phi(x)$ and of its source density $J(x)$ will not be fixed in this paper. If $\Phi(x)$ is a gauge field some gauge-fixing constraint must of course be considered, but here we will be concerned with solutions from the wave equation only. This usually would imply on a larger solution space but, as we will see, for a discrete field this is not the case. 

Eqs. (\ref{fitotfif}),(\ref{eq1}) and (\ref{eq2}) imply that for a discrete field, the Eq. (\ref{gfeq}) becomes 
\be 
\l{gffeq}
{\cal O}(x,\tau,g,\nabla,s)\Phi_{f}(x,\tau)=J(x,\tau),
\ee
which does not represent a new postulated equation; it is the same operator of Eq. (\ref{gfeq}), in a new notation in order to take explicit advantage of the field constraint (\ref{f}), applied on a discrete field. $g$ is the metric tensor induced \cite{BJP} by the constraint (\ref{f}) in the embedded (3+1) manifold:
\be
\l{gd}
g_{\mu\nu}=\eta_{\mu\nu}+f^{2}f_{\mu}f_{\nu},
\ee
and its inverse, $g_{\mu\nu}g^{\nu\alpha}=\delta^{\alpha}_{\mu}$,
\be
\l{gu}
g^{\mu\nu}=\eta^{\mu\nu}-\frac{f^{2}f^{\mu}f^{\nu}}{1+(f^2)^2},
\ee
with $f^2=-1$ for a massive field, and $f^2=0$ for a massless one (for which then $g_{\mu\nu}=\eta_{\mu\nu}$).

The Eq. (\ref{gffeq}) is solved, using a Green's function, by 
\be
\label{sgf}
\Phi_{f}(x,\tau_{x})=\int d^{4}yd\tau_{y}\; G_{f}(x-y,\tau_{x}-\tau_{y})\;J(y,\tau_{y}),
\ee
where the sub-indices in $\tau$ specify the respective events $x$ and $y$, and $G_{f}(x-y,\tau_{x}-\tau_{y})$ is a solution of
\be
\label{gfe} 
{\cal O}(x,\tau,g,\nabla,s)G_{f}(x-y,\tau_{x}-\tau_{y})=\delta^{4}(x-y)\delta(\tau_{x}-\tau_{y}).
\ee

\section{Causality and dynamics}

As already mentioned, classical electrodynamics uses explicitly both constraints (\ref{1}) and (\ref{f}) but in a way that is not entirely consistent for not recognizing the second constraint as being distinct from the first one, that they carry distinct informations. 
In general Eqs. (\ref{1}) and (\ref{f}) are just two kinematical constraints on a field propagation but the second one acquires a dynamical content when $\Delta x$ describes the  spacetime separation between two physical objects like a source and its field, as discussed in \cite{hep-th/9708096}.  
$\Delta x=x-z(\tau)$ for a field emitted by a point charge at $z(\tau)$ on its worldline, which is taken, using Eq. (\ref{1}), as parameterized by  its $\tau$.  In the limit of $x$ tending to $z(\tau)$ both $\Delta\tau$ and $\Delta x$ go to zero. Nothing changes with respect to the constraint (\ref{1}) but there is a crux change with respect to the constraint (\ref{f}) because $f$ is not well defined in this limit. It requires a more careful analysis than the one usually found in the literature, which just does not consider the fact that at the vertex of a cone its tangent vector is not defined. This is the origin of inconsistencies of classical electrodynamics which are usually glossed over with the assumption that at a such limit quantum electrodynamics should take over. 
A more appropriate treatment \cite{hep-th/9610028} explains out the inconsistency and implies on a finite field with a finite energy content. There is no infinity. 

For a massless field the restriction (\ref{f}) is reduced to $f.(x-z(\tau))=0$ and this implies that the event $x$, where the field is being observed, and the charge retarded position  $z(\tau)$ must belong to a same null line $f$.  It is not necessary to explicitly distinguish a generic $\tau$ from a $\tau$ at a retarded position, as the situations considered in this paper, from now on, will always refer to the last one. 

More information can be extracted from this constraint as $\partial_{\mu}f.(x-z)=0$ implies on 
\be
\l{fv} 
f.u=-1,
\ee
where $u=({\vec u},u_{4})=\frac{dz}{d\tau}$ is the source four-velocity. This relation  may be seen as a covariant normalization of the time component of $f$ to 1 in the charge rest-frame at its retarded time, 
\be
\l{f4}
f^{4}{\Big |}_{{\vec u}=0}=|{\vec f}|{\Big |}_{{\vec u}=0}=1.
\ee
From Eq. (\ref{fv}), with $a^{\mu}=\frac{du^{\mu}}{d\tau}$, we get 
\be
\l{dA0}
a.f=0,
\ee
a constraint 
between the direction $f$ along which the signal is emitted (absorbed) and the instantaneous change in the charge state of motion at the retarded (advanced) time. It implies that
\be
\l{a4}
a_{4}=\frac{{\vec a}.{\vec f}}{f_{4}},
\ee
whereas $a.u\equiv0$ leads to
$a_{4}=\frac{{\vec a}.{\vec u}}{u_{4}},$
and so we have that in the charge instantaneous rest frame at the emission (absorption) time ${\vec a}$ and ${\vec f}$ are orthogonal vectors,
\be
\l{af}
{\vec a}.{\vec f}{\Big |}_{{\vec u}=0}=0.
\ee
The constraint (\ref{dA0}) has been obtained here on very generic grounds of causality, without reference to any specific interaction, which makes of it a universal relation, valid for all kinds of fields and sources. It is remarkable that this same behaviour is predicted to hold for all fundamental (strong, weak, electromagnetic and gravitational) interactions. For the electromagnetic field this is an old well known and experimentally confirmed fact that takes, in the standard formalism of continuous fields, the whole apparatus of Maxwell's theory to be demonstrated \cite{Rorhlichp112}. Its experimental confirmation is a direct validation of extended causality. This is discussed, in terms of discrete fields, in \cite{hep-th/9911233,paperIII}.

\section{The discrete Green's function.} 

In this section we will be interested on the discrete version of the Green's function for the wave equation 
\be
\label{kgwe}
(\eta^{\mu\nu}\partial_{\mu}\partial_{\nu}-m^2)\Phi(x)=J(x),
\ee
associated to the Klein-Gordon operator for a massive field on a Minkowski manifold and that is solved by the continuous Green's function $G(x)$, given in the Eq. (\ref{Gd}) below. Applied to a discrete field, the Eq. (\ref{kgwe}) should become
\be
\label{dkgwe}
(g^{\mu\nu}\nabla_{\mu}\nabla_{\nu}-m^2)\Phi_{f}(x,\tau)=J(x,\tau),
\ee 
as $det\;g=const.$ 
But this equation does not have a consistent solution\cite{ecwpd} unless $f^2=0$ and the mass term be dropped from it so that it is reduced to
\be
\label{dkgwesm}
\eta^{\mu\nu}\nabla_{\mu}\nabla_{\nu}\Phi_{f}(x,\tau)=J(x,\tau).
\ee 
Otherwise, it would produce\cite{ecwpd} a non-propagating discrete field which would be a violation of the Lorentz symmetry, so that no physical object can be described by the Eq. (\ref{dkgwe}). But interestingly, this does not mean that Eq. (\ref{dkgwesm}) is applicable only to massless fields; as we will see, it applies to both massive and massless fields. The mass term and the timelike velocity  are hidden behind a non-constant $\tau$ for not spoiling the field conformal symmetry.

The Eq. (\ref{dkgwesm}) is solved\cite{ecwpd} to give:
\be
\label{pr99}
G_{f}(x,\tau)=\frac{1}{2}ab\varepsilon\theta(a\tau)\theta(b{\bar f}.x)\delta(\tau+ f.x),\;\;\;\;{\vec x}_{{\hbox{\tiny T}}}=0,
\ee
or, equivalently by
\be
\label{pr9}
G_{f}(x,\tau)=\frac{1}{2}ab\varepsilon\theta(a\tau)\theta(bt)\delta(\tau+ f.x),\;\;\;\;{\vec x}_{{\hbox{\tiny T}}}=0,
\ee
where $a,b,\varepsilon=\pm1,$ as the signs of $\tau$, $t$ and $f^4$, respectively. 
 They are restricted by $ab\varepsilon=1$. $\theta (x)$ is the Heaviside function, $\theta(x\ge0)=1$ and $\theta(x<0)=0.$  For $f^{\mu}=({\vec f}, f^{4})$, ${\bar f}$ is defined by ${\bar f}^{\mu}=(-{\vec f}, f^{4}).$ 
The subscript ${\hbox {\tiny T}}$ stands for transversal with respect to ${\vec f}$: $${\vec f}.{\vec x}_{{\hbox {\tiny T}}}=0.$$

The meaning of the append ${\vec x}_{{\hbox{\tiny T}}}=0$ is that the $G_{f}$-defining Eq. (\ref{gfe}) is effectively replaced by
\be
\l{Gm}
\eta^{\mu\nu}\nabla_{\mu}\nabla_{\nu} G_{f}(x,\tau)=\delta(x_{{\hbox{\tiny L}}})\delta(t)\delta(\tau),\;\;\;\;{\vec x}_{{\hbox{\tiny T}}}=0,
\ee
where the subscript ${\hbox {\tiny L}}$ stands for longitudinal with respect to ${\vec f}$:
$$x_{{\hbox {\tiny L}}}=\frac{{\vec f}.{\vec x}}{|{\vec f}|},$$ so that $$\int_{{\vec x}_{{\hbox{\tiny T}}}=0}d^5x\eta^{\mu\nu}\nabla_{\mu}\nabla_{\nu} G_{f}(x,\tau)=1.$$ The append ${\vec x}_{{\hbox{\tiny T}}}=0$ restricts the integration domain so that Eq. (\ref{sgf}) is equivalent to
\be
\label{sgfr}
\Phi_{f}(x,\tau_{x})=\int dy_{{\hbox{\tiny L}}}dt_{y}d\tau_{y}\; G_{f}(x-y,\tau_{x}-\tau_{y})\;J(y,\tau_{y}),\;\;\;\;{\vec x}_{{\hbox{\tiny T}}}={\vec y}_{{\hbox{\tiny T}}}.
\ee
Under the integration sign this append can, of course, be replaced by a factor $\delta^{2}({\vec x}_{{\hbox{\tiny T}}}-{\vec y}_{{\hbox{\tiny T}}})$ but never on equations  (\ref{pr99}) and (\ref{pr9}). 

The most obvious difference between $G_{f}(x,\tau)$ and $G(x)$, and consequently, between $\Phi_{f}$ and $\Phi$, is the absence of singularity. The discrete field propagates without changing its amplitude. Such a so great difference between  two fields generated by a same source is closely associated to the distinct topologies of their respective supports. A cone is not a complete manifold in contradistinction to any of its generators. Conceptually important is that the singularity $r=0$ (in $G(x)$ of Eq. (\ref{Gd}) below), which gives origin to an infinite self energy for the continuous field, is not, as will be clear, a consequence of self interactions but just from its mathematical representation.  

\begin{center}

a) $(1+1)$ effective dynamics

\end{center}

Before going to discuss the meaning of the solution (\ref{pr9}) let us explore the fact that it
 does not depend on ${\vec x}_{\hbox {\tiny T}}$.  
\be
\l{Gt}
\frac{\partial}{\partial x_{{\hbox{\tiny T}}}}G_{f}(x,\tau)=0.
\ee
The interaction propagated by $G_{f}$ is blind to ${\vec x}_{{\hbox{\tiny{T}}}}$. Anything at the transversal dimensions is not affected by and do not contribute to the interaction described by $G_{f}(x,\tau)$. Although we are working with a field formalism defined on a (3+1)-spacetime, with respect to its dynamics the spacetime is effectively reduced to a (1+1)-spacetime, without any breaking of the explicit Lorentz covariance. 

Distinct uncharged fields emitted by neighbouring point-sources do not see each other; each one of them can be treated as an independent single entity. 
This suggests the interpretation of $\Phi_{f}$ as a physical point object since its propagation does not depend on ${\vec x}_{\hbox {\tiny T}}$, or in other words, on anything outside $f$. This interpretation is reinforced in papers II and III where the energy-momentum conservation of discrete fields with specific tensoriality is shown.

Eq. (\ref{Gt}) has a further consequence that a discrete field has a necessarily discrete point source: the origin of the signal represented by $G_{f}(x,\tau)$ must be a point at the intersection (in the past) of the straight line $f$ with $J(x,\tau)$, and this must be an isolated event. In an extended source this event would have neighbouring events that could not be just ignored because they  would induce a continuity not consistent with Eq. (\ref{Gt}).  This is then a formalism of discrete fields, discrete sources and discrete interactions; apparent continuity being just a matter of scale. Discrete here means pointlike, structureless. There is a complete symmetry between fields and sources. They are all discrete and obeying to the same causality constraints (\ref{1}) and (\ref{f}); they are all discrete fields. This is a relevant symmetry because what would be an apparent weakness, the restriction to discrete sources, turns into an unifying principle valid for all fundamental fields (fermions and bosons, in the words of a quantum context). For the sake of simplicity, as we focus rather on $\Phi_{f}$ as a discrete field, we will omit the discrete-field character of $J(x,\tau)$ \cite{hep-th/9708096} treating it just as a standard pointlike object. So, from now on, although not explicitly said, the discrete-field properties to be discussed are shared also by the field sources as they are discrete fields too.

In this formalism where $\tau$ is treated  as a fifth independent coordinate the point source-density of a discrete field must carry an additional constraint expressing the causal relationship between two events, say, $x$ and $z$:
\be    
\l{jtfd}
J(x,\tau_{x}=\tau_{z})=j(\tau)\delta^{(4)}(x-z(\tau))=j(\tau)\delta^{3}({\vec x}-{\vec z(\tau)})\delta(t_{x}-t_{z}(\tau)).
\ee
That $\tau_{x}=\tau_{z}$ on the LHS is a consequence of the Eq. (\ref{1}) which is used only afterwards, so that Eq. (\ref{jtfd}) may also be written as $J(x,\tau)-j(\tau)\delta^3(x)\delta(\tau).$  The standard current $J(x)$ is then related to $J(x,\tau)$ by

\be
\l{JtoJ}
J(x)=\int d\tau J(x,\tau).
\ee
$j(\tau)$ defines the tensorial or spinorial character of both $J(x,\tau)$ and $J(x)$.

Sometimes it may be  convenient to replace the wave equation (\ref{dkgwesm}) by 
\be
\label{wefm}
\eta^{\mu\nu}\nabla_{\mu}\nabla_{\nu}\Phi_{f}(x-z,\tau_{x}-\tau_{z})=J_{[f]}(x-z,\tau_{x}-\tau_{z}),\;\;\;\;{\vec x}_{{\hbox{\tiny T}}}={\vec z}_{{\hbox{\tiny T}}},
\ee
where
\be
\l{dsf}
J_{[f]}(x-z,\tau_{x}-\tau_{z})=j(\tau)\delta(\tau_{x}-\tau_{z})\delta(t_{x}-t_{z})\delta(x_{\hbox{\tiny L}}-z_{\hbox{\tiny L}}),,\;\;\;\;{\vec x}_{{\hbox{\tiny T}}}={\vec z}_{{\hbox{\tiny T}}}
\ee
is the source density stripped of its explicit ${\vec x}_{\hbox {\tiny T}}$-dependence, which is, by the way, irrelevant because  Eq. (\ref{1}) implies on
\be
\l{1a}
(t_{x}-t_{z})^2=(x_{{\hbox{\tiny L}}} -z_{{\hbox{\tiny L}}})^2+(x_{{\hbox{\tiny T}_{1}}} -z_{{\hbox{\tiny T}_{1}}})^2+(x_{{\hbox{\tiny T}_{2}}} -z_{{\hbox{\tiny T}_{2}}})^2+(\tau_{x}-\tau_{z})^2,
\ee
and the deltas in the definition (\ref{dsf}) imply that
\be
\l{xt12}
\delta(\tau_{x}-\tau_{z})\delta(t_{x}-t_{z})\delta(x_{\hbox{\tiny L}}-z_{\hbox{\tiny L}})\Longrightarrow\cases{x_{{\hbox{\tiny T}
_{1}}} =z_{{\hbox{\tiny T}_{1}}}&\cr
	&\cr
	x_{{\hbox{\tiny T}_{2}}} =z_{{\hbox{\tiny T}_{2}}},\cr}
\ee

where we have used the notation
\be
x:=({\vec x},t_{x})=(x_{{\hbox{\tiny L}}},x_{{\hbox{\tiny T}_{1}}},x_{{\hbox{\tiny T}_{2}}},t_{x}).
\ee

More generally, $t=0$ in $t^2=\tau^2+x^2$ defines the hypercone vertex and so it implies on $\tau=x=0.$
The following effective identities are useful:
\be
\l{iden}
\delta(t)\delta(\tau){\Big|}_{\tau^2+x^2=0}=\delta(t)\delta(x_{i}){\Big|}_{\tau^2+x^2=0}=\delta(\tau)\delta(x_{i}){\Big|}_{\tau^2+x^2=0}
\ee
where $x_{i}$ stands for any space component of ${\vec x}$. In their demonstrations there is a subtle passage $$\delta(\sqrt{-x^2}\;)=\delta(\sqrt{x^2}\;)$$ which is, of course, valid only for $x=0$, that is, at the cone vertex.

So Eqs. (\ref{sgf}) and (\ref{sgfr}) are equivalent to
\be
\l{phif}
\Phi_{f}(x,\tau_{x})=\int dy_{{\hbox{\tiny L}}}dt_{y}d\tau_{y}G_{f}(x-y,\tau_{x}-\tau_{y})J_{[f]}(y,\tau_{y}),
\ee
where $\Phi_{f}(x,\tau_{x})$ and $J_{[f]}(y,\tau_{y})$ respectively means $\Phi_{f}(x_{\hbox{\tiny L}},{\vec x}_{\hbox{\tiny T}}={\vec z}_{\hbox{\tiny T}},t_{x},\tau_{x}=\tau_{z})$ and $J_{[f]}(y-z,\tau_{y}=\tau_{z}).$  From now on, ${\Big|}_{f}$ will be an indication of $${\Big|}_{{{\vec x}_{\hbox{\tiny T}}=0}\atop{{\tau+f.x=0}\atop{\tau^2+ x^2=0}}}.$$
In contraposition to $J$, $J_{[f]}$ is just a formal definition; it is explicitly dependent on $f$ and should not be confused with the support of $J$ which is not being shown in $j(\tau)$.
If we use Eqs. (\ref{jtfd}) and (\ref{pr9}) in Eq. (\ref{sgf}) we obtain
\be
\Phi_{f}(x,\tau)=\frac{1}{2}\theta(a\tau)\theta(bt)j(\tau){\bigg|}_{f},
\ee
and then we can see that the field $\Phi_{f}$ is completely determined  by its (discrete) source current $j(x)$. This exposes again the field-source symmetry in a discrete field approach. A discrete field is equivalent to a unidimensional current so that charge conservation ($\nabla.J=0$) and the Lorentz gauge condition of a gauge field ($\nabla.A=0$) are both inherent properties of a discrete field, a consequence of Eq. (\ref{dA0}). This is being discussed in paper III
.  Hereon we will focus on the properties, physical meaning and consequences of its $G_{f}$ as they are universal, valid for all fundamental interactions.
\begin{center}
b) Massive and massless fields
\end{center}
If $\Delta\tau=0$ then $\Delta x$ is a four vector collinear to $f$, and $G_{f}(x,\tau)$ describes a massless point signal propagating along the null direction $f$. 
If $\Delta\tau\neq0$ one has a massive point signal propagating along a $\Delta x$ not collinear to $f$. There is then a timelike four vector $v$ such that 
\be
\Delta\tau+v.\Delta x=0,\;\;\;{\hbox{or}}\;\;\;v=\frac{\Delta x}{\Delta\tau},
\ee
with $v.f=v^2=-1$
which is compatible with Eq. (\ref{f}) because 
\be
\l{nfv}
(f-v).\Delta x:=n.\Delta x=0
\ee
defines 
\be
\l{n}
n=f-v
\ee
 with $$n.v=0$$ so that
\be
\l{vf}
n.f=1,\;\;\;n^2=1.
\ee
$n$ is a spacelike four vector that in the rest-frame of the signal (the one where $\Delta x=(\Delta t,{\vec0})$) it is given by $n=(0,{\vec n})$ with $|{\vec n}|=1$. The implication of  Eq. (\ref{nfv}) is that for $\Delta\tau\ne0$,  Eq. (\ref{pr9}) describes the propagation of a massive signal on a timelike $\Delta x$ through its  projection on a null direction $f$: $v.\Delta x=f.\Delta x.$

 The point is that the constraints (\ref{1}) and (\ref{f}) are valid for both cases, $\Delta\tau=0$ and $\Delta\tau\ne0.$  There is nothing in Eq. (\ref{pr9}) that fixes $\Delta\tau$  or makes any reference to $v$ or to any mass. It is indeed remarkable that it describes the propagation of both massive and massless fields and not only of the massless one as one could expect from the Eq. (\ref{dkgwe}). A Lagrangian  for a massive discrete field cannot have an explicit mass term nor any reference to its velocity $v$, and must be written in terms of its projection on a lightcone, that is, in terms of $f$ and not of $v$. It describes a discrete massive field as it were massless; a non-constant proper-time is its telltale. So it is not surprising that it be invariant under conformal transformations, as it carries no fixed scale. This is like what happens in gauge field theories where the Lagrangian cannot have an explicit mass term for preserving the gauge symmetry; here the preserved symmetry is the conformal one. 

As there is nothing in Eq. (\ref{pr9}) that assures that $\Delta\tau=0,$ this condition for a massless field must be fixed by an afterwards explicit specification 
\be
\l{ts2}
\Phi_{f}(x-z,\tau_{x}=\tau_{z})=\Phi_{f}(x-z,\tau_{x}-\tau_{z}){\Big |}_{\tau_{x}=\tau_{z}}=\int d\tau_{z}\delta(\tau_{x}-\tau_{z})\Phi_{f}(x-z,\tau_{x}-\tau_{z}).
\ee
\begin{center}
c) Creation and annihilation processes
\end{center}
$\Phi_{f}$  is just a point on a straight line propagating with the four-velocity $f$, if massless, or $v$ if massive.  It is the reduction of the field support from a lightcone to a lightcone generator that makes the discrete field to be just a point in the phase space.
$f$ and ${\bar f}$ are two opposing generators of a same lightcone; they are associated, respectively, to the $b=+1$ and to the $b=-1$ solutions and, therefore, to the processes of creation and annihilation of a discrete field. See the Figure 2. 

\hspace{-3cm}
\vglue-3cm

\begin{minipage}[]{7.0cm}
\hspace{-1cm}
\parbox[t]{5.0cm}{
\begin{figure}
\vglue-3cm
\epsfxsize=400pt
\epsfbox{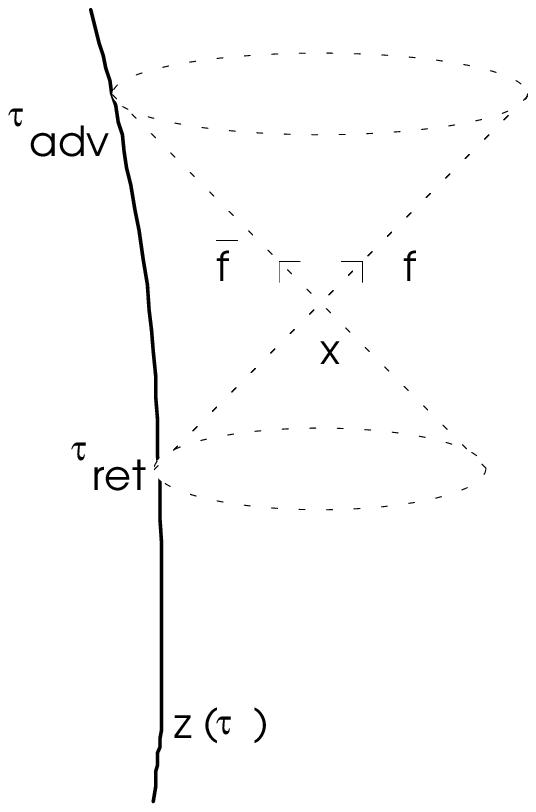}
\end{figure}}
\end{minipage}\hfill
\vglue-7cm

\hglue7.0cm

\begin{minipage}[]{7.0cm}\hglue12.0cm
\begin{figure}
\hglue7.0cm
\l{f3a}
\parbox[t]{5.0cm}{\vglue-6cm\hglue7.0cm{
\caption[Creation and annihilation of discrete fields]{Creation and annihilation of particles: The discrete solutions from the wave equation as creation and annihilation of particle-like fields. There are two discrete fields, represented by arrows, at the point x: one, created at $\tau_{ret}$, has propagated to x on the lightcone generator f; the other one propagating on the lightcone generator ${\bar f}$ from x towards the charge world line where it will be annihilated at $\tau_{adv}$. The charge on the world line $z(\tau)$ is a source for the first and a sink for the second.}}}
\end{figure}
\end{minipage}

\vglue-1cm
The arrows in the Figure 2 represent the propagating fields.
Observe that there is no backwards propagation in time implying that there is no advanced solution; the creation and the annihilation solutions are both retarded.
For $b=+1$ or $t>0$, $G_{f}(x,\tau)$ describes a point signal emitted by the charge  at $\tau_{ret},$ and that has propagated to $x$ along the fibre $f$ of the future lightcone of $z(\tau_{ret})$;  for $b=-1$ or $t<0,$  $G_{f}(x,\tau)$ describes a point signal that is propagating  along the fibre $\bar{f}$ of the future lightcone of $x$ towards the point $z(\tau_{adv})$ where it will be absorbed (annihilated) by the charge. 
 The only difference between the $(b=+1)$ and the $(b=-1)$ solutions is that $J$ is the source for the first and the sink for the second. Nothing else.
Observe the differences from the interpretation of the corresponding standard solutions. There is no advanced, causality violating solution here. These two solutions correspond to creation and annihilation of discrete fields, exactly like well known processes of creation and annihilation of particles in relativistic quantum field theory. The difficulty with a continuous classical field is that one has point sources that creates it but no point sink that annihilates it. It cannot be symmetric as it would require a, possibly infinite, continuous distribution of absorbers. 
\begin{center}
c) Checking the discrete solution
\end{center}
Since we left the derivation of Eq. (\ref{pr9}) for the reference \cite{ecwpd} it may be instructive to verify that it is indeed a solution of Eq. (\ref{Gm}). 
As
\be
\l{tfx}
\nabla(\tau+f.x)\equiv 0,
\ee
\be
\eta^{\mu\nu}\nabla_{\mu}\nabla_{\nu} \theta(b\tau)=\nabla^{\nu}(-bf_{\nu}\delta(b\tau))=-f^{2}\delta'(\tau)=o,
\ee
and 
\be
\eta^{\mu\nu}\nabla_{\mu}\nabla_{\nu}\theta(-b{\bar f}.x)=b{\bar f}_{\nu}\nabla^{\nu}\delta(-b{\bar f}.x)=f^{2}\delta'({\bar f}.x)=o,
\ee
because of Eq. (\ref{fd}) and $f^2={\bar f}^2=0$, 
we find that
$$
\eta^{\mu\nu}\nabla_{\mu}\nabla_{\nu} G_{f}(x)=\delta(\tau+f.x)\nabla\theta(b\tau).\nabla\theta(-b{\bar f}.x)=-f.{\bar f}\delta(\tau)\delta({\bar f}.x)\delta(\tau+f.x)=$$
$$=-(f_{4}^2+|{\vec f}|^2)\delta(\tau)\delta(f_{4}t-|{\vec f}|x_{\hbox{\tiny L}})\delta(f_{4}t+|{\vec f}|x_{\hbox{\tiny L}})=2f_{4}^{2}\delta(\tau)\delta(2f_{4}t)\delta(|{\vec f}|x_{\hbox{\tiny L}})=\delta(\tau)\delta(t)\delta(x_{\hbox{\tiny L}}).
$$ 

\section{Conformal symmetry}
We show now that the causality constraints (\ref{1}) and (\ref{f}) imply on $G_{f}(\Delta x,\Delta\tau)$ invariant under conformal transformations. First we ask which transformations
\be
\l{cs}
\Delta x^{\mu}\rightarrow \Delta x^{\mu}+\delta_{{\hbox{\tiny T}}}^{\mu}(\Delta x)
\ee
 of four-vectors $\Delta x$ on a Minkowski manifold leave invariant the constraint (\ref{1}) as it induces a transformation $$\Delta\tau\delta_{{\hbox{\tiny T}}}(\Delta\tau)+\Delta x.\delta_{{\hbox{\tiny T}}}(\Delta x)=0$$ so that, according to Eq. (\ref{ffrac})
\be
\l{deltaTtau}
\delta_{{\hbox{\tiny T}}}(\Delta\tau)=\cases{-v_{\mu}\delta^{\mu}_{{\hbox{\tiny T}}}(\Delta x)& if $\Delta\tau\ne0$,\cr
&\cr
-f_{\mu}\delta^{\mu}_{{\hbox{\tiny T}}}(\Delta x)& if $\Delta\tau=0.$}
\ee
For $\Delta\tau=0$ the answer is already known \cite{Cunningham}. Let us find the symmetry of $\Delta\tau+f.\Delta x$ with $f$ being a null vector irrespective of wether $\Delta\tau$ is null or not. For $\Delta\tau=0$ $f$ and $\Delta x$ are  parallel vectors, $\Delta x^2=0,$  and the symmetry of $f.\Delta x$ is the same of $\Delta x^2$, i.e. the conformal one. For $\Delta\tau\neq0$ we have from Eq. (\ref{deltaTtau}) 
\be
\l{46p}
\delta_{{\hbox{\tiny T}}}(\Delta\tau+f.\Delta x)=(f-v)_{\mu}\delta^{\mu}_{{\hbox{\tiny T}}}(\Delta x)+\Delta x_{\mu}\delta^{\mu}_{{\hbox{\tiny T}}}(f).
\ee
Then using Eq. (\ref{n}) and (\ref{nfv}) we get  $$\delta_{{\hbox{\tiny T}}}(\Delta\tau+f.\Delta x)=-n_{\mu}\delta^{\mu}_{{\hbox{\tiny T}}}(\Delta x)+\Delta x_{\mu}\delta^{\mu}_{{\hbox{\tiny T}}}(v-n)=$$
\be
=-\delta_{{\hbox{\tiny T}}}(n.\Delta x)+\Delta x_{\mu}\delta^{\mu}_{{\hbox{\tiny T}}}(v)=\Delta x_{\mu}\delta^{\mu}_{{\hbox{\tiny T}}}(v),
\ee
as $n.\Delta x=n.v\Delta\tau=0$ by construction.
Then
\be
\delta_{{\hbox{\tiny T}}}(\Delta\tau+f.\Delta x)=\Delta x_{\mu}\delta^{\mu}_{{\hbox{\tiny T}}}(\frac{\Delta x}{\Delta\tau})=\frac{\Delta x_{\mu}}{\Delta\tau}[\delta^{\mu}_{{\hbox{\tiny T}}}(\Delta x)-\frac{\Delta x^{\mu}}{\Delta\tau}\delta_{{\hbox{\tiny T}}}(\Delta\tau)]=(1+\frac{\Delta x^2}{\Delta\tau^2})\frac{\Delta x_{\nu}}{\Delta\tau}\delta^{\nu}_{{\hbox{\tiny T}}}(\Delta x)\equiv0.
\ee
Therefore, the constraints (\ref{1}) and (\ref{f}) have the same symmetries. We can see from Eq. (\ref{46p}) that $\Delta\tau+v.\Delta x$ has also these 
same symmetries but it would not produce{\footnote{A lightike $f$ in Eq. (\ref{dkgwesm}) produces a propagator with one pole instead of the two of a 
timelike one. This makes the whole difference.}} a solution to the wave equation \cite{ecwpd}.

Considering the manifestly explicit Lorentz covariance of Eq. (\ref{pr9}) it is enough to verify the invariance under
\be
\l{cases}
\cases{\delta_{D}^{\mu}(\Delta x)=\Delta x^{\mu};&\cr
&\cr
\delta_{C_{\nu}}^{\mu}(\Delta x)=2\Delta x^{\mu}\Delta x_{\nu}-\eta^{\mu}_{\nu}\Delta x^{\rho}\Delta x_{\rho}.\cr}
\ee
Let us do it explicitly for the second one (the other one is just as easy). 
\be
\delta_{C_{\nu}} [G_{f}(\Delta x,\Delta\tau)]=\delta(\Delta\tau+f.\Delta x)[\theta(\Delta t)\delta _{C_{\nu}}(\theta(\Delta\tau))+\theta(\Delta\tau)
\delta _{C_{\nu}}(\theta(\Delta t))]=0
\ee
because
\be
\delta _{C_{\nu}}(\theta(\Delta\tau))=-\delta(\Delta\tau)f_{\mu}\delta^{\mu} _{C_{\nu}}(\Delta x)=
-\delta(\Delta\tau)(2f.\Delta x\Delta x_{\nu}+f_{\nu}\Delta\tau^2)=2\Delta x_{\nu}\Delta\tau\delta(\Delta\tau)=0,
\ee
and
\be
\delta _{C_{\nu}}(\theta(\Delta t))=\delta(\Delta t)\delta _{C_{\nu}}(\Delta t)=\delta(\Delta t)(2\Delta t\Delta x_{\nu}-\eta^{4}_{\nu}\Delta\tau^2)
=-\eta^{4}_{\nu}\delta(\Delta t)\Delta\tau^2=0,
\ee 
where the last passage is a consequence of Eq. (\ref{1}) or, explicitly, $\Delta t^{2}=\Delta\tau^{2}+r^{2},\;\;r=|{\vec x}|$, so that $\Delta t=0$ implies 
necessarily on $\Delta\tau^{2}=0.$ The two solutions, with either $f$ or ${\bar f}$, in Eq. (\ref{pr99}) correspond to the right and left movers of the (1+1)
-physics \cite{1+1} and they have the same symmetry.

This conformal invariance is not much of a surprise after seing the last section but it is nonetheless interesting that massive and massless fields in a 
(3+1) space-time have, equally, a conformal symmetry. Its price is of keeping hidden all masses and timelike velocities.

\section{Retrieving the usual frame work}

In this section we discuss the passage from the discrete to the standard formalism of continuous fields.
\begin{center}
a) Massless fields
\end{center}

Here we prove the following connection between $G(x)$, solution of Eq. (\ref{kgwe}), and $G_{f}(x,\tau)$ of Eq. (\ref{pr9}): 
\be
\l{gg}
G(x,\tau)=\frac{1}{2\pi}\int d^4f\;\delta(f^2)G_{f}(x,\tau),
\ee
with
\be
\l{df4}
d^{4}f=df_{4}\;|{\vec f}|^{2}\;d|{\vec f}|\;d^{2}\Omega_{f},
\ee
\be
\l{delta2}
\delta(f^{2})=\frac{1}{2|{\vec f}|}\{\delta(f^{4}-|{\vec f}|)+\delta(f^{4}+|{\vec f}|)\},
\ee
and 
\be
\l{gg1}
G(x)=\int d\tau G(x,\tau)\delta(\tau)\delta(\tau-\sqrt{-x^2}).
\ee

We write $f.x=f_{4}t+r|{\vec f}|\cos\theta_{f},$ where $r=|{\vec x}|$ and the angle $\theta_{f}$ is defined by
\be
\l{theta}
{\vec f}.{\vec x}:= r|{\vec f}|\cos \theta_{f},
\ee
for a fixed $ x$. Then we have from Eqs. (\ref{pr9}) and (\ref{gg}-\ref{delta2}), after using 
 $\varepsilon=\frac{f^4}{|f^4|}=-\frac{f_4}{|f^4|}$ in the integration on $f_{4}$, 
\be
\l{gg2}
G(x,\tau)=-\frac{ab\theta(a\tau)\theta(bt)}{8\pi}\int |{\vec f}| d|{\vec f}|d^2\Omega_{f}{\Big\{}\delta[\tau+|{\vec 
f}|(r\cos\theta_{f}+t)]-\delta[\tau+|{\vec f}|(r\cos\theta_{f}-t)]{\Big \}}
\ee
and then

\be
\l{22}
G(x,\tau)=\frac{ab\theta(a\tau)\theta(bt)}{4}\int_{-1}^{1}d\cos\theta_{f} \{\frac{\tau}{(r\cos\theta_{f}+t)|r\cos\theta_{f}+t|}-\frac{\tau}{(r\cos\theta_{f}-t)|r\cos\theta_{f}-t|}\}.
\ee
The constraint (\ref{1}), $t^2=\tau^2+r^2$, implies that $|t|\ge r$, so that
\be
\l{tmr}
|r\cos\theta_{f}\pm t|=\frac{t}{|t|}(t\pm r\cos\theta_{f})=b\theta(bt)(t\pm r\cos\theta_{f})
\ee
and Eq. (\ref{22}) then becomes
\be
\l{qq}
G(x,\tau)=G(t,r,\tau)=\frac{a\theta(a\tau)}{4}\int_{-1}^{1}d\cos\theta_{f}\;{\big\{}\frac{\tau}{(t+r\cos\theta_{f})^{2}}-\frac{\tau}{(t-r\cos\theta_{f})^{2}}{\big\}}=
\ee
$$=\frac{a\theta(a\tau)}{2}\int_{-1}^{1}
\frac{\tau}{(t+r\cos\theta_{f})^{2}}d\cos\theta_{f}.
$$
Therefore, we have
\be
\l{G}
G(x,\tau)=G(t^2-r^2,\tau)=a\theta(a\tau)\frac{\tau}{(t^2-r^2)},
\ee

which, with the use of Eq. (\ref{1}) gives
\be
\l{qqq}
G(t^2-r^2,\tau){\Big |}_{\tau=0}=\cases{0,& for $|t|-r\ne0$;\cr
                   \infty,& for $|t|-r=0$.\cr}
\ee
The RHS of  Eq. (\ref{G}) represents a spherical signal propagating with the velocity of light, as expected. So, with
\be
\l{dtau}
\delta(\tau)=\delta(\sqrt{t^2-r^2}\;)=\frac{|\tau|}{r}[\delta(t-r)+\delta(t+r)]=\frac{a\tau}{r}[\delta(t-r)+\delta(t+r)],
\ee
in Eq. (\ref{gg1}) we have that
$$G(x)=\int d\tau\frac{a\tau}{t^2-r^2}\delta(\tau)\delta(\tau-\sqrt{t^2-r^2})$$
\be
\l{Gd}
G(r,t)=\frac{1}{r}[\delta(t-r)+\delta(t+r)]=2\delta(t^2-r^2),
\ee
and then from Eqs. (\ref{JtoJ}), (\ref{sgf}), (\ref{gg}) and (\ref{gg1}) 
\be
\label{s1s}
\Phi(x)=\frac{1}{2\pi}\int d^{4}f\;\delta(f^{2})\Phi(x,\tau_{x}=\tau_{z})_{f}.
\ee
$\Phi$ represents rather the smearing of $\Phi_{f}$ over the lightcone. 
For the emitted field $(f^{4}=|{\vec f}|)$ in the source instantaneous rest frame at the emission time $(f^{4}=1$, according to Eq. (\ref{f4})), the Eq. (\ref{s1s}) can be written as
\be
\label{s}
\Phi(x,\tau)=\frac{1}{4\pi}\int d^{2}\Omega_{f}\Phi(x,\tau)_{f},
\ee
where the integral represents the sum over all directions of ${\vec f}$ on  a lightcone. $4\pi$ is a normalization factor 
\be
\label{sss}
\Phi(x,\tau)=\frac{\int_{\Omega_{f}} d^{2}\Omega_{f}\Phi(x,\tau)_{f}}{\int_{\Omega_{f}} d^{2}\Omega_{f}}.
\ee

\begin{center}
b) Massive fields
\end{center}
For the case of a massive field{\footnote{Using the generic metric tensor (\ref{gd}) we could deal with both massive and massless cases at once.} we need to 
retrieve $v$ from the data $(\Delta x,\Delta\tau,f)$ and change the integration variable from $f$ to $p$ with $p=mv$. We start from
\be
\l{mgg1}
G(x,\tau)=\frac{m}{2\pi^2}\int d^4f\;d\tau\;\delta(f^2)G_{f}(x,\tau)\delta(f.v+1),
\ee
with $\delta(f.v+1)$ instead of $\delta(\tau)$ of the massless case, for making, according to Eqs. (\ref{n}) and (\ref{vf}), the connection between $f$ and $v$, which implies also that $$\delta(f^2)=\delta(v^2+1)=m^2\delta(m^2+p^2)$$ and $$\delta(\tau+f.x)=\delta(\tau+v.x)=m\delta(m\tau+p.x).$$ The extra $\frac{m}{\pi}$ factor in Eq. (\ref{mgg1}) is for keeping the normalization.
Then
\be
\l{mgg2}
G(x,\tau)=\frac{ab}{8\pi^{3}}\theta(a\tau)\theta(bt)\int d^4f\;df_{5}\varepsilon\;m^4\delta(p^2+m^2)\delta(\frac{f.p}{m}+1)e^{if_{5}(m\tau+p.x)}, 
\ee
where we have used an integral form for delta in the Green's function (\ref{mgg1}). Now a change of integration variables $(f,f_{5})\Rightarrow(p,f_{5})$ with
\be
f:=\frac{p}{mf_{5}}
\ee
so that 
\be
\frac{\partial(f,f_{5})}{\partial(p,f_{5})}=\frac{1}{(f_{5})^4m^4},
\ee
leads to
\be
G(x,\tau)=\frac{ab\theta(a\tau)\theta(bt)}{8\pi^{3}}\int\frac{ d^4p}{|f_{5}|^3}\;df_{5}\varepsilon\;\delta(p^2+m^2)\delta(\frac{p^2}{m^2}+f_{5})e^{if_{5}(m\tau+p.x)}.
\ee
The integration on $f_{5}$ gives
$$G(x,\tau)=e^{im\tau}\frac{ab\theta(a\tau)\theta(bt)}{8\pi^{3}}\int d^4p\varepsilon\;\delta(p^2+m^2)e^{ip.x}=$$
$$=e^{im\tau}\frac{ab\theta(a\tau)\theta(bt)}{8\pi^{3}}\int \frac{d^4p}{2|p_{4}|}[\delta(p_{4}-\sqrt{({\vec p})^2+m^2})-\delta(p_{4}+\sqrt{({\vec p})^2+m^2})]e^{ip.x},$$
where we made use of $\varepsilon=-\frac{p_4}{|p_4|}$. So the Cauchy's theorem can be used for writing
\be
\l{mgg4}
G(x,\tau)=e^{im\tau}\frac{a\theta(a\tau)}{8\pi^{3}}\frac{1}{2\pi}\int \frac{d^4p}{ p^2+m^2\pm i\epsilon}e^{ip.x},
\ee
which reproduces the standard \cite{BD} Klein-Gordon Green's function $G(x)$ of a massive field with an extra factor
\be
\l{factor}
G(x,\tau)=a\theta(a\tau)e^{im\tau}G(x).
\ee
$G(x)$ is a solution to
\be
\l{standwe}
(\eta^{\mu\nu}\partial_{\mu}\partial_{\nu}+m^2)G(x)=\delta^4(x).
\ee
So we can replace Eq. (\ref{s1s}) with a more generic one valid for both massive and massless fields 
\be
\label{s1sg}
a\theta(a\tau)e^{im\tau}\Phi(x)=\frac{1}{2\pi}\int_{hc} d^{4}f\;\Phi_{f}(x,\tau_{x}),
\ee
where the integration is over a hypercone (hc), a restriction that stands for an integrand factor: $\delta(f^2)\delta(\tau)$ for a massless field, and for $\frac{m}{\pi}\delta(v^2+1)\delta(f.v+1)$ for a massive one. Of course, for a massless field $a\theta(a\tau)e^{im\tau}=1$.

\begin{center}
c) The wave equation.
\end{center}

 Starting with Eqs. (\ref{sgf}) and (\ref{dkgwesm}) with its LHS expanded we have 
\be
\frac{1}{2\pi}\int_{hc} d^{4}f\;\eta^{\mu\nu}(\partial_{\mu}\partial_{\nu}-2f_{\mu}\partial_{\nu}\partial_{\tau_{x}}+f_{\mu}f_{\nu}\partial^2_{\tau_{x}}){\Big\{}\Phi_{f}(x,\tau_{x})-\int d^5yG_{f}(x-y,\tau_{x}-\tau_{y})J(y,\tau_{y}){\Big\}}=0.
\ee
Under the integration over the $f$ degrees of freedom the terms linear in $f_{\mu}$ do not contribute 
 as $$\int_{hc} d^{4}ff^{\mu}\partial_{\mu}\partial_{\tau}G_{f}(x,\tau)=0$$
$$\int_{hc} d^{4}ff^{\mu}\partial_{\mu}\partial_{\tau}\Phi_{f}(x,\tau)=0$$ because $G_{f}(x,\tau)$ and $\Phi_{f}(x,\tau)$ are even functions of $f$, as we can see from Eqs. (\ref{ffrac}), (\ref{fmu}) and (\ref{pr9}), i.e. $f\rightarrow-f$, with a fixed $\tau$, implies on $x\rightarrow-x$ which leaves $G_{f}$ and, therefore, $\Phi_{f}$ invariant. For a massless field $f^2=0$ and so, in this case, the terms quadratic in $f$ do not contribute,  whereas for a massive field, $f^2=-1$, they can be moved out from the integration sign, 
\be
(\eta^{\mu\nu}\partial_{\mu}\partial_{\nu}+f^2\partial^2_{\tau_{x}})\frac{1}{2\pi}\int_{hc} d^{4}f\;{\Big\{}\Phi_{f}(x,\tau_{x})-\int d^5yG_{f}(x-y,\tau_{x}-\tau_{y})J(y,\tau_{y}){\Big\}}=0,
\ee
so that with Eqs. (\ref{factor}) and (\ref{s1sg}) we have
\be
(\eta^{\mu\nu}\partial_{\mu}\partial_{\nu}+f^2\partial^2_{\tau_{x}}){\Big\{}a\theta(a\Delta\tau)e^{im\Delta\tau}\Phi(x)-\int d^4y\;a\theta(a\Delta\tau)e^{im\Delta\tau}G(x-y)J(y){\Big\}}=0,
\ee
where the remaining $f^2$ stands for zero and -1 for the massless and massive fields, respectively, and we have used Eq. (\ref{JtoJ}). Then, after doing the $\tau$-derivatives we end up with two equations
\be
\Phi(x)=\int d^4yG(x-y)J(y),
\ee
from the imaginary part, and using Eq. (\ref{standwe})
\be
(\eta^{\mu\nu}\partial_{\mu}\partial_{\nu}+m^2)\Phi(x)=J(x),
\ee
from the real part, for both the massive and the massless fields.

The $f$-integration erases in the wave equation the effects of the constraint (\ref{f}) on the field. So, the standard continuous formalism is retrieved from this discrete $f$-formalism with $\Phi(x)$ as the average of $\Phi_{f}(x)$ in the sense of Eq. (\ref{s1s}), and its wave equation as the average of Eq. (\ref{dkgwesm}), for both, the massive and the massless fields. 

\section{Mass and symmetry breaking}
It may sound strange, at a first sight, that we have both massiveness and conformal symmetry on a same field. Conformal symmetry requires and implies the 
inexistence of any scale whereas a mass or a timelike velocity represents one ($p^2=-m^2$, $v^2=-1$). Although we are not considering here any 
particular tensor/spinor character for the discrete field it is worthwhile to extend this discussion also to the case of the chiral symmetry of a fermionic 
massive discrete field. In the following considerations a timelike four-velocity and a non-longitudinal (with respect to velocity) spin component have  
the same role on the broken chiral and conformal symmetries, respectively. 
respectively, the chiral and the conformal symmetries.

The point is that a discrete field $\Phi_{f}$ does not properly describe a physical object but its projection, or rather one of its two projections on the 
lightcone.  They are orthogonal projections made through the plane defined by the t-axis and $v$, the four-velocity of the proper physical discrete field. 
See the Figure 3.

\vglue13cm

\hglue-3cm

\begin{minipage}[]{7.0cm}

\parbox[t]{5.0cm}{

\begin{figure}

\vglue-13cm
\epsfxsize=350pt
\epsfbox{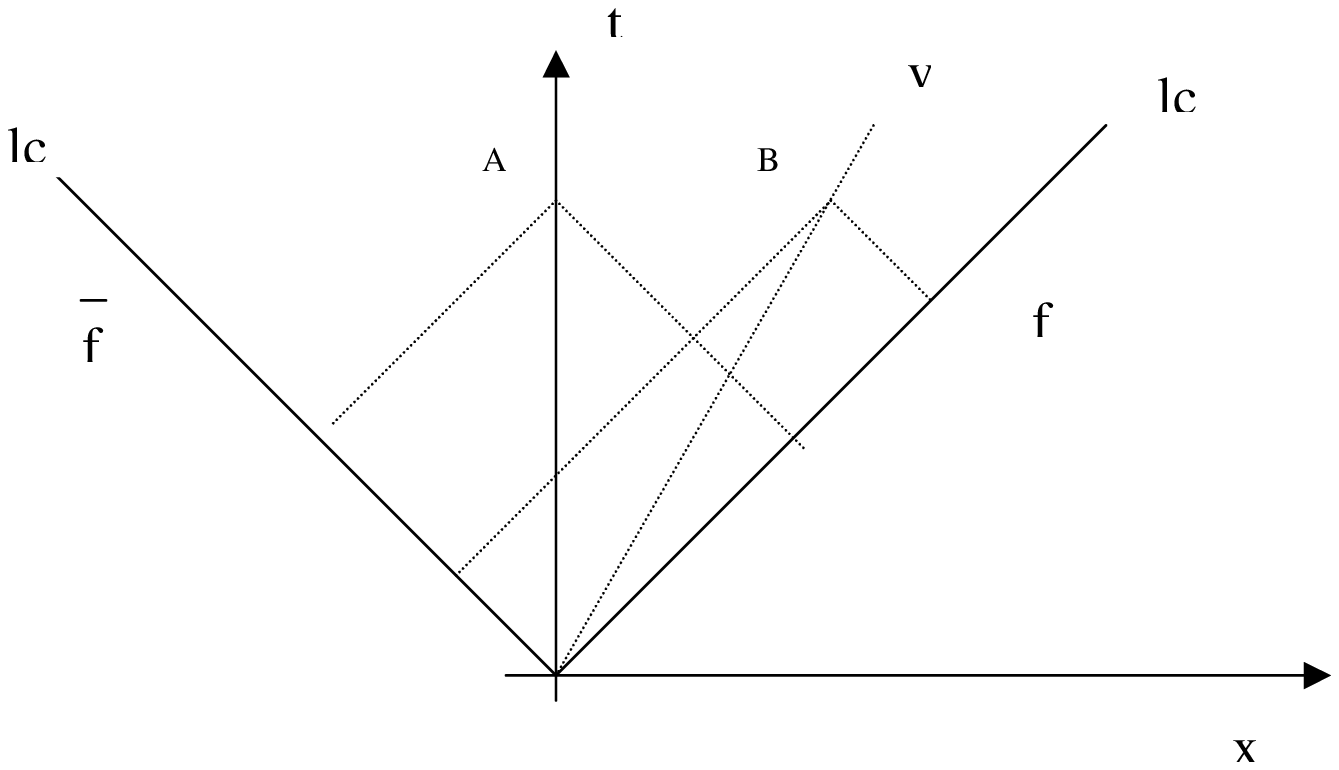}

\end{figure}}

\end{minipage}\hfill

\vglue-5cm
\hglue-1.0cm

\begin{minipage}[]{7.0cm}

\begin{figure}

\hglue9.3cm

\parbox[t]{5.0cm}{\vglue-10cm

\caption{Mass and symmetry breaking. The conformal symmetry of massive discrete fields is understood as the symmetry of their projections $\Phi_{f}$ 
and $\Phi_{{\bar f}}$ on the lightcone (lc). A discrete field (A) in its rest-frame has equal components on a basis $\{|f>\}$ isomorphic to the set 
of lightcone generators. This composition is, of course, frame dependent. A massive discrete field has support on a generator $v$ (its four-velocity) 
of a hypercone totally inside the lightcone lc.}}

\end{figure}

\end{minipage}

\vglue-1cm

\vglue-3cm

A massless physical object is on the lightcone and so it coincides with its own projection. A massive one always has two projections, one 
along $f$ and the other along ${\bar f}$. $f=({\vec f}, f^4)$ and ${\bar f}=({-\vec f}, f^4)$. Both propagating forward on time, as it happens to any discrete 
field.

Its Hamiltonian must be a function of both projections $$H=H(\Phi_{f},\Phi_{{\bar f}}).$$ For a massive field $H$ is diagonal on a basis $\{|f>\}$ defined by 
its isomorphism with the set of lightcone generators $f.$ For a massive field $H$ is diagonal on a basis $\{|v>\}$, defined by the set of 
generators $v$ of a hypercone that is 
totally inside the lightcone. See the Figure 3. $$|v>=\alpha|f>+\beta|{\bar f}>.$$  In its rest-frame the two projections have a same amplitude: 
$\alpha =\beta.$  Its mass spectrum is defined by the set of eigenvalues of $H$. The chiral and the conformal symmetry are assured for the projections by 
$f^2={\bar f}^2=0$ but not for the massive discrete field because $v^2=2\alpha\beta{\bar f}.f\neq0.$ Therefore the change of basis 
\be
\l{cb}
\{|f>\}\Rightarrow\{|v>\}
\ee
exposes the broken symmetries and introduce a scale in the theory because it changes the description from the projections to the proper massive discrete field. 

In the passage from discrete to continuous fields in Eq. (\ref{mgg1}) the role of $\delta(f.v+1)$ is of establishing the proportion of $f$ and of ${\bar f}$ in 
the 
composition of $v$. Of course, we can have symmetry breaking with the change of basis (\ref{cb}) and without leaving the discrete formalism. See 
\cite{hadrons}, 
for an example.

\section{Conclusions}

This work throws some light on the meaning and origin of continuous fields, their symmetries and singularities through the introduction of discrete fields, a 
mixed concept of particles (discreteness, pointlikeness) and fields (differentiability and the superposition principle). Although we have left any further 
discussion about the physical meaning of a discrete field for the companion papers II and III we have seen enough to say that with $\Phi_{f}$  the continuous 
$\Phi$ becomes its effective average on the lightcone and that the need of boundary conditions, the lose connection between $\Phi$ and its sources, its gauge 
freedom and singularity can all be better understood. 

The relationship between a discrete field and the standard continuous field is similar to the one between thermodynamics and statistical mechanics. The 
meaning and the origins of thermodynamical variables (P, T, S, etc) are illuminated by statistical mechanics from the knowledge of basic structural elements 
unknown to thermodynamics. Similarly, properties of  a continuous field (its problems, symmetries and singularities) are better understood from its relation 
to its discrete field. Thermodynamics is an effective description in terms of average valued properties of the more basic structures, the molecules,  
considered in statistical mechanics; likewisely, the various forms of field theory (general relativity, statistical mechanics and quantum mechanics included) 
are retrieved from their respective formulation in terms of discrete fields as effective average-valued descriptions. Statistical mechanics does not change 
the status of any of the four laws of thermodynamics, it just put them in a deeper perspective. Analogously, the validity and  the fundamental character of 
the field  equations remain with the discrete fields and they are equally seen from a deeper perspective. Although not being a fundamental 
theory thermodynamics is successfully used when this is more convenient than using the known basic molecular structures. The same goes for the continuous 
and the discrete fields.


\begin{thebibliography}{100}
\bibitem {hep-th/9610028} M. M. de Souza, J. of Phys. A: Math. Gen. 30 (1997)6565-6585. Hep-th/9610028.
\bibitem{Rorhlich}F. Rorhlich, {\it Classical Charged Particles},  Reading, Mass. (1965).
\bibitem{Jackson} D. Jackson {\it Classical Electrodynamics},2nd ed., chaps. 14 and 17,
John Wiley {\&} Sons, New York, NY(1975).
\bibitem{Ternov} A.A. Sokolov, I. M. Ternov, {\it Radiation from relativistic electrons}, A.I.P, New York, N. Y, 1986.
\bibitem{Teitelboim} C. Teitelboim; D. Villaroel; Ch.G.Van Weert, Rev. del
Nuovo Cim., vol 3, N.9,(1980).
\bibitem{Rowe} E.G.P.Rowe, Phys. Rev. D, 12, 1576(1975); 18, 3639(1978);  Nuovo Cim, B73, 226(1983).
\bibitem {Lozada} A. Lozada, J. Math. Phys., 30,1713(1989).
\bibitem {gr-qc/9801040} M. M. de Souza, Robson N. Silveira,  Class. \& and Quantum Gravity, vol 16, 619(1999). Gr-qc/9801040
\bibitem {paperII} M.M. de Souza, {\it Discrete scalar fields and general relativity}. Hep-th/0006250. 
\bibitem {gr-qc/9903071} M. M. de Souza, Robson N. Silveira,{\it Gauss vs Coulomb and the cosmological mass deficit problem.} Gr-qc/9903071
\bibitem{hep-th/9708096}M. M. de Souza, {\it Dynamics and causality constraints.} Hep-th/0006214. 
\bibitem{paperIII}M.M. de Souza, {\it Photons as discrete fields and the wave-particle duality}. In preparation.
\bibitem {hep-th/9911233} M.M. de Souza,{\it {Discrete gauge fields.}} Hep-th/9911233
\bibitem {Berkov}See for example A.V.Berkov, Yu.P. Nikitin, I.L.Rosental< Fortschrite der Physic  29,303(1981), and the references therein.
\bibitem {string}J. Polchinski, {\it String Theory. Vol. I}, Cambridge Univ. Press, Cambridge(1998).
\bibitem{Schouten}J.A. Schouten and J. Haantjes, Physica1, 869(1934).
\bibitem{Barut}A.O. Barut and R.B. Haugen, Ann. Phys. 71, 519(1972); Nuovo Cimento 18A, 511(1973).
\bibitem {KK}See, for example, {\it An introduction to Kaluza-Klein theories}, H.C.Lee(ed.), World Scientific, Singapore(1984) and the references therein.
\bibitem {Dirac2times}P.A.M.Dirac, Ann. Math. 37(1936)429
\bibitem {Bars}I. Bahrs, Phys. Rev. D59(1999) 045019
\bibitem{tau}Fanchi, J. R., Found. Phys 23, 487(1993);Gaioli, F. H., Garcia-Alvarez, E. T., Int. J. Phys, 36, 2391(1997).
\bibitem{Rorhlichp112}See, for example, the reference \cite{Rorhlich} at page 112.
\bibitem {hep-th/9610145} M. M. de Souza, {\it Classical Fields and the Quantum concept.}Hep-th/9610145.
\bibitem {BJP} M. M. de Souza, Braz. J. of Phys., vol 28, n. 3, 250(1999).
\bibitem{Cunningham} E. Cunningham, Proc. Math. Soc. London 8, 77(1909); H. Bateman, {\it ibid.} 8, 1686(1910); G. Mack, A. Salam, Ann. Phys. 53, 174(1969); 
H.A. Kastrup, Phys. Rev. 150, 1189(1964).
\bibitem{1+1}D. Boyanovsky and C. M. Naon, Riv. Nuovo Cim., vol 13, 2,1(1990).
\bibitem{ecwpd}Theory of discrete fields. In preparation.
\bibitem{BD}J.D. Bjorken, S.D. Drell, {\it Relativistic Quantum Mechanics}, McGraw-Hill(1964).
\bibitem{hadrons} M. M. de Souza, {\it The origin of mass and the electroweak mass spectrum without Higgs} in {\it Hadron Physics-94}, V. Herscovitz, C.A.Z. 
Vasconcellos, E. Ferreira (eds.) 265, World Scientific, Singapure(1994).
\end{thebibliography}
\end{document}